\documentclass[journal=langd5,manuscript=article]{achemso}

\usepackage{chemformula} 
\usepackage[T1]{fontenc} 

\newcommand{\RR}[0]{\boldsymbol{R}}
\newcommand{\cc}[0]{\mathrm{c}}
\newcommand{\ee}[0]{\mathrm{e}}
\newcommand{\sss}[0]{\mathrm{s}}
\newcommand{\tot}[0]{\mathrm{tot}}

\author{Jaime Agudo-Canalejo}
\affiliation[MPIDS]
{Department of Living Matter Physics, Max Planck Institute for Dynamics and Self-Organization, D-37077 G\"ottingen, Germany}
\author{Pierre Illien}
\affiliation[Sorbonne]
{Sorbonne Universit\'e, CNRS, Laboratoire Physicochimie des Electrolytes et Nanosyst\`emes Interfaciaux (PHENIX), UMR  8234, 4 place Jussieu, 75005 Paris, France}
\author{Ramin Golestanian}
\email{ramin.golestanian@ds.mpg.de}
\affiliation[MPIDS]
{Department of Living Matter Physics, Max Planck Institute for Dynamics and Self-Organization, D-37077 G\"ottingen, Germany}
\alsoaffiliation[Oxford]
{Rudolf Peierls Centre for Theoretical Physics, University of Oxford, Oxford OX1 3PU, United Kingdom}

\title[An \textsf{achemso} demo]
  {Comment on ``Relative Diffusivities of Bound and Unbound Protein Can Control
  	Chemotactic Directionality''}


\begin{document}

In a recent study \cite{mandal} published in \emph{Langmuir}, Mandal and Sen claim to propose a ``new'' kinetic model to analyze the directional movement of enzyme molecules in response to a gradient of their substrate, with the supposedly new prediction that net movement occurs up the substrate gradient when the diffusivity of the substrate-bound enzyme is lower than that of the unbound enzyme, and movement down the substrate gradient when the
diffusivity of the substrate-bound enzyme is higher than that of the unbound enzyme. With the present Comment, we would like to point out that the exact same result and prediction (with an identical derivation) was already obtained by us as one of the central results in Ref.~\citenum{agudo2018phoresis}, whose Abstract indeed reads that we found ``a new type of [chemotactic] mechanism due to binding-induced changes in the diffusion coefficient of the enzyme'' which ``points toward lower substrate concentration if the substrate enhances enzyme diffusion and toward higher substrate concentration if the substrate inhibits enzyme diffusion.''

This would not require any additional explanation had Mandal and Sen been unaware of our work, as rediscovery of known phenomena is a common-enough occurrence in science. However, Mandal and Sen repeatedly cite and discuss Ref.~\citenum{agudo2018phoresis}, widely misrepresenting it and falsely claiming (in order of appearance) that our approach:
\begin{itemize}
\item ``[assumes] that the effective diffusivity of the protein is the weighted average of the diffusivity of free and bound protein.''
\item ``[does not make] a distinction between the mass ﬂuxes of the free and the bound protein''	
\item ``is in contrast with [their approach]''
\item ``fails to recognize the gradients of the free and bound protein that are created because of the presence of the ligand gradient''
\item ``seriously underestimates the chemotaxis of the protein when there is no initial gradient of the protein in the system''
\item ``[ignores] two terms that are incorporated in [their eq. 6]''
\end{itemize}
As we show below, the derivation and, consequently, the central result (eq. 6) of Mandal and Sen are \emph{identical} to those in Ref.~\citenum{agudo2018phoresis}, and therefore all their claims listed above are unjustified.

We begin by noting that our derivation in Ref.~\citenum{agudo2018phoresis} starts from a fully-stochastic description of the enzyme and substrate molecules, and furthermore includes the possibility of hydrodynamic and non-specific enzyme-substrate interactions. After making a mean field approximation for the substrate concentration, it is shown that the combination of non-specific and hydrodynamic interactions results in an additional, \emph{phoretic} mechanism for chemotaxis that is not taken into account by Mandal and Sen. The results of Mandal and Sen are therefore a special case of ours (corresponding to setting $\boldsymbol{v}_\mathrm{e}=\boldsymbol{v}_\mathrm{c}=0$ in eqs.~6, 7, and 15 of Ref.~\citenum{agudo2018phoresis}). In what follows we discuss only this special case.

The equivalence in notation between our work \cite{agudo2018phoresis} and Mandal and Sen's \cite{mandal} is summarized in Table~\ref{tbl:notation}, while the equivalence between equations, which for the purpose of this Comment we will number (I--IV), is summarized in Table~\ref{tbl:eqs}. By simply contrasting the versions of (I), (II), and (III) in Ref.~\citenum{mandal} with those of Ref.~\citenum{agudo2018phoresis} it is obvious that they are manifestly identical. Because (IV), which is the central result in both works, is directly derived from (I--III) in exactly the same way in both works, it must necessarily be identical in both works as well. Any illusory perception of Mandal and Sen's results being different to ours must thus come from the way that (IV) is presented in each case.

\begin{table}
	\caption{Equivalence table for notations}
	\label{tbl:notation}
	\begin{tabular}{lll}
		\hline
		Meaning  & Ref.~\citenum{mandal} & Ref.~\citenum{agudo2018phoresis}  \\
		\hline
		Free enzyme concentration   & $c_\mathrm{A}$ & $\rho_\mathrm{e}$   \\
		Enzyme-substrate complex concentration & $c_\mathrm{AB}$ & $\rho_\mathrm{c}$  \\
		Total enzyme concentration & $c_\mathrm{A}^\mathrm{T}=c_\mathrm{A}+c_\mathrm{AB}$ & $\rho_\mathrm{e}^\mathrm{tot} = \rho_\mathrm{e} + \rho_\mathrm{c}$ \\
		Substrate concentration  & $c_\mathrm{B}$ & $\rho_\mathrm{s}$  \\
		Free enzyme diffusion coefficient & $D_\mathrm{A}$ & $D_\mathrm{e}$ \\
		Enzyme-substrate complex diffusion coefficient & $D_\mathrm{AB}$ & $D_\mathrm{c}$ \\
		Substrate binding rate & $k_1$ & $k_\mathrm{on}$ \\
		Substrate unbinding rate & $k_{-1}$ & $k_\mathrm{off}$ \\
		Dissociation constant & $K_\mathrm{d} = k_{-1}/k_{1}$ & $K = k_\mathrm{off}/k_\mathrm{on}$ \\
		\hline
	\end{tabular}
\end{table}

\begin{table}
	\caption{Equivalence table for equations}
	\label{tbl:eqs}
	\begin{tabular}{llll}
		\hline
		~ & Meaning & Ref.~\citenum{mandal} & Ref.~\citenum{agudo2018phoresis}  \\
		\hline
		(I) & Evolution of free enzyme concentration & Eq.~2 & Eq.~6   \\
		(II) & Evolution of enzyme-substrate complex concentration & Eq.~3 & Eq.~7  \\
		(III) & Assumption of instantaneous local binding equilibrium & Eq.~5  & Eq.~11  \\
		(IV) & Evolution of total enzyme concentration & Eq.~6 & Eqs.~13--16 \\
		\hline
	\end{tabular}
\end{table}

In Ref.~\citenum{agudo2018phoresis}, we presented (IV) as 
\begin{equation}
\partial_t \rho_{\ee}^{\tot}(\RR;t) = \nabla \cdot \left[ D(\RR)\cdot\nabla \rho_{\ee}^{\tot} -  \boldsymbol{V}_\text{bi}(\RR) \rho_{\ee}^{\tot} \right],
\label{eq:evol}
\end{equation}
with the definition of an effective, substrate-concentration-dependent diffusion coefficient
\begin{eqnarray}
D(\RR) &=& D_{\ee} + (D_{\cc} - D_{\ee})   \frac{\rho_{\sss}(\RR)}{K + \rho_{\sss}(\RR)},
\label{eq:D}
\end{eqnarray}
and a binding-induced chemotactic velocity
\begin{eqnarray}
\boldsymbol{V}_\mathrm{bi}(\RR) &=& - (D_{\cc} - D_{\ee}) \nabla \left(  \frac{\rho_{\sss}(\RR)}{K + \rho_{\sss}(\RR)} \right).
\label{eq:Vbi}
\end{eqnarray}
Eq.~\ref{eq:evol} here has the advantage of being written in a canonical form, with the total enzyme flux being cleanly split into a Fickian diffusion flux $-D(\RR)\cdot\nabla \rho_{\ee}^{\tot}$, and an advective, chemotactic flux $\boldsymbol{V}_\text{bi}(\RR) \rho_{\ee}^{\tot}$. In particular, in the absence of substrate gradients, the latter chemotactic flux vanishes and one is left with Fickian diffusion only.

The result for (IV) of Mandal and Sen \cite{mandal} is identical to this one, just presented in a non-canonical form that mixes diffusive and chemotactic fluxes. Indeed, plugging in the expressions for $D(\RR)$ and $\boldsymbol{V}_\mathrm{bi}(\RR)$ into eq.~\ref{eq:evol} above and rearranging the gradient terms, one can trivially rewrite eq.~\ref{eq:evol} as
\begin{equation}
\partial_t \rho_{\ee}^{\tot}(\RR;t) = D_\ee \nabla^2 \rho_{\ee}^{\tot} +  (D_{\cc} - D_{\ee}) \nabla^2 \left(  \frac{\rho_{\ee}^{\tot} \rho_{\sss}}{K + \rho_{\sss}} \right),
\label{eq:evol2}
\end{equation}
which now makes explicit that Mandal and Sen's result is identical to ours. This form of the equation is not particularly transparent, however, as the second term also contributes to diffusion, and is non-zero even if the substrate concentration is uniform in space.

For completeness, we note that there are other instructive ways in which this same evolution equation can be written. For example, in Ref.~\citenum{agudo2018enhanced}, we pointed out that it can also be equivalently rewritten as 
\begin{equation}
	\partial_t \rho_{\ee}^{\tot}(\RR;t) = \nabla^2 \left[ D(\RR)  \rho_{\ee}^{\tot} \right],
	\label{eq:evol3}
\end{equation}
with $D(\RR)$ given by eq.~\ref{eq:D} above, which implies that, in the absence of enzyme sources and sinks, and in the presence of an externally-maintained substrate gradient, the enzyme concentration will reach a zero-flux stationary state with $\rho_{\ee}^{\tot}(\RR) \propto 1/D(\RR)$, i.e.~will accumulate in regions where the effective diffusion coefficient is lowest.

In summary, Mandal and Sen \cite{mandal} seem to have misunderstood the results in Ref.~\citenum{agudo2018phoresis}, which are identical to theirs (although Ref.~\citenum{agudo2018phoresis} additionally includes the possibility of phoresis arising from nonspecific and hydrodynamic interactions). While, in light of this, the central message of Mandal and Sen (i.e.~that ``relative diffusivities of bound and unbound protein can control chemotactic directionality'' as per the title) is not new, we would like to note that their work does bring some new and interesting aspects to the literature. In particular, (i) the inclusion of the catalytic step (with catalytic rate $k_2$ in Ref.~\citenum{mandal}, $k_\mathrm{cat}$ in Ref.~\citenum{agudo2018phoresis}) which was neglected in Ref.~\citenum{agudo2018phoresis} (by considering the limit $k_\mathrm{cat} \ll k_\mathrm{off}$); as well as (ii) their numerical simulation of the transient kinetics in a setting that mimics a microfluidics experiment, which moreover helps in ascertaining the range of validity of the instantaneous local binding equilibrium assumption. 

To finish, we note that, since the publication of Ref.~\citenum{agudo2018phoresis}, there have been some further developments of the idea of chemotaxis resulting from binding-induced changes in diffusivity. In Refs.~\citenum{adeleke2019chemical} and \citenum{agudo2020diffusion}, it was shown that the same mechanism operates for non-rigid enzymes or proteins that undergo shape fluctuations, in which case the binding-induced changes in diffusion that cause chemotaxis can come not only from changes in the average shape of the protein, but also in the magnitude of its shape fluctuations. In Ref.~\citenum{agudo2020diffusion}, it was explicitly shown that the competition between phoretic and binding-induced mechanisms for chemotaxis can lead to accumulation or depletion of enzymes not just in regions of highest or lowest substrate concentration, but also in regions with an intermediate, tunable critical substrate concentration. Lastly, in Ref.~\citenum{agudo2020cooperatively}, it was shown that a similar mechanism for chemotaxis due to changes in diffusivity operates in the case of oligomeric proteins that can reversibly associate and dissociate into monomers. Such oligomeric proteins spontaneously accumulate in regions in which the oligomeric (slowly-diffusing) form is most stable, in a process termed ``stabilitaxis''.

%
%

%
%
%

\bibliography{biblio}

\end{document}